\documentclass{iau} 
\usepackage{graphicx}

\title[Globular clusters in the era of precision astrometry] 
{Globular clusters in the era of precision astrometry }

\author[P. Bianchini]   
{Paolo Bianchini$^1$
 }

\affiliation{$^1$ Universit\'{e} de Strasbourg, CNRS, Observatoire Astronomique de Strasbourg, UMR 7550, F-67000 Strasbourg, France \\email: {\tt paolo.bianchini@astro.unistra.fr}}

\pubyear{2019}
\volume{351}  
\jname{Star Clusters: From the Milky Way to the Early Universe}
\editors{A. Bragaglia, M.B. Davies, A. Sills \& E. Vesperini, eds.}
\begin{document}

\maketitle

\begin{abstract}
The study of the kinematics of globular clusters (GCs) offers the possibility of unveiling their long term evolution and uncovering their yet unknown formation mechanism. \textit{Gaia} DR2 has strongly revitalized this field and enabled the exploration of the 6D phase-space properties of Milky Way GCs, thanks to precision astrometry. However, to fully leverage on the power of precision astrometry, a thorough investigations of the data is required. In this contribution, we show that the study of the mean radial proper motion profiles of GCs offers an ideal benchmark to assess the presence of systematics in crowded fields. Our work demonstrates that systematics in \textit{Gaia} DR2 for the closest 14 GCs are below the random measurement errors, reaching a precision of $\sim0.015$ mas yr$^{-1}$ for mean proper motion measurements. Finally, through the analysis of the tangential component of proper motions, we report the detection of internal rotation in a sample of $\sim50$ GCs, and outline the implications of the presence of angular momentum for the formation mechanism of proto-GC. This result gives the first taste of the unparalleled power of \textit{Gaia} DR2 for GCs science, in preparation for the subsequent data releases.

\end{abstract}

\firstsection 
\section{Introduction}
The traditional picture of globular clusters (GCs) as simple stellar systems is being radically revolutionized by the synergistic efforts of theoretical advances and precision astrometry led by \textit{Gaia} DR2 (\cite{Gaia2018}). We now know that GCs harbour rich kinematic features, such as the presence of internal rotation (e.g. \cite{Bianchini2013,Kamann2018}) and complex stellar populations (e.g. \cite{Milone2018}). These elements strongly puzzle our understanding of their formation in the early universe.

In this context, large kinematic data sets are the fundamental tool to address the open questions on GCs formation and evolution. However, until recently, only a handful of GCs have been studied kinematically, often only via small samples of line-of-sight velocity measurements (\cite{Zocchi2012}) or \textit{Hubble Space Telescope (HST)} proper motions for a few targeted GCs (\cite{Libralato2018,Bellini2017,Watkins2015}). The advent of \textit{Gaia} DR2 has dramatically changed the field, enabling a full 6D view of Milky Way GCs throughout their spatial extent. Current kinematic studies are now based on a factor of $\approx50$ more stars with exquisite accuracy measurements (compared to the pre-Gaia era, see e.g. \cite{Zocchi2012} and \cite{Bianchini2018} for \textit{Gaia} DR2), for nearly all GCs (\cite{Baumgardt2018}), enabling access to their internal kinematics in the regime $<1$ km s$^{-1}$ (typically $<0.05$~mas~yr$^{-1}$).

However, to fully leverage on the power of precision astrometry, a thorough investigation of the data is first required. The challenge consists of disentangling complex physical dynamical processes from subtle data systematics at the sub-km s$^{-1}$ regime.
In this work, as a first step, we show how to exploit GCs as a benchmark to quantify subtle systematics in crowded fields, in particular using their mean radial proper motion profiles (Section \ref{sec:radial}). Subsequently, we investigate the presence of internal rotation in a sample of ~50 GCs, analyzing the tangential component of proper motions (Section \ref{sec:tangential}), and outline the implications of the presence of angular momentum for the formation mechanism and evolution of proto-GC. The main results of this work are presented in \cite[Bianchini et al. (2018)]{Bianchini2018} and \cite[Bianchini et al. (2019)]{Bianchini2019}.

\section{Radial component of proper motions}
\label{sec:radial}
\begin{figure}[b]
\begin{center}
 \includegraphics[width=0.9\textwidth]{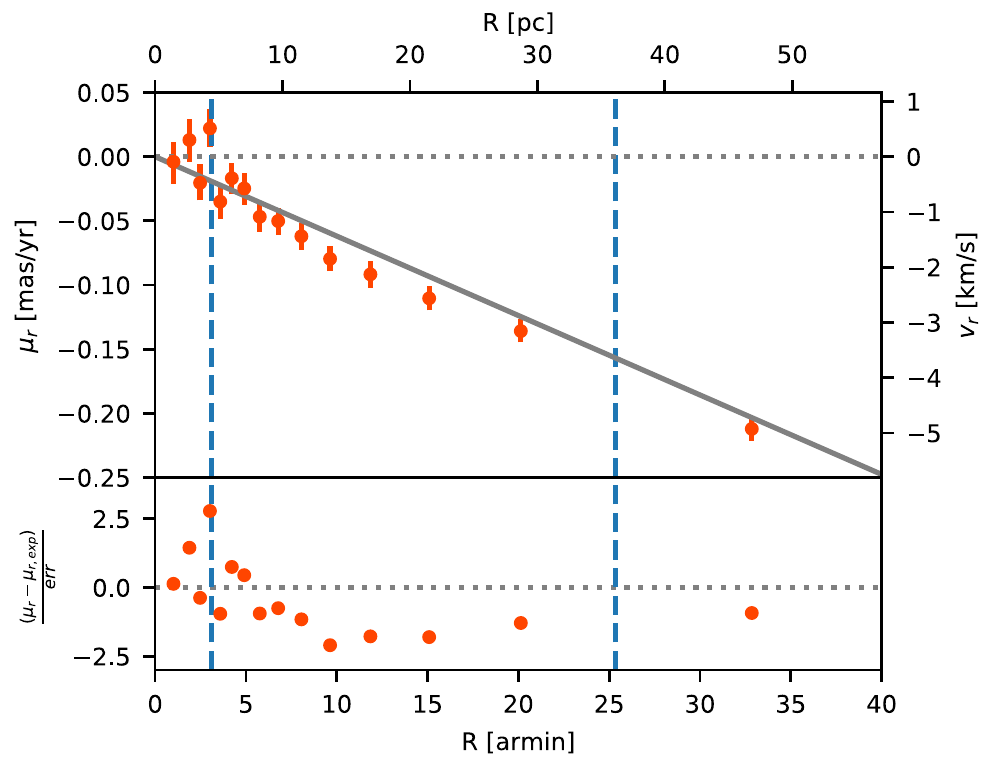} 
 \caption{\textbf{Mean profile of the radial proper motion component of NGC 3201}. The cluster shows a significant perspective contraction due to its high receding motion along the line-of-sight ($V_\mathrm{los}=494.3$ km s$^{-1}$, \cite[Harris 1996]{Harris1996}). The bottom panel illustrates that the mean discrepancy between the observed mean radial proper motion and the expected one (eq. \ref{eq:expansion}) is within $\lesssim2$-sigma. This implies that data systematics are within the statistical error ($\lesssim0.014$~mas~yr$^{-1}$ / 0.32 km s$^{-1}$).}
   \label{fig1}
\end{center}
\end{figure}

The analysis of the radial component of proper motions $\mu_r(R)$ reveals the presence (or absence) of expansion/contraction in GCs. If GCs are in dynamical equilibrium, standard dynamical processes (such as core collapse) are not able to imprint a measurable expansion/contraction signature. However, perspective effects due to their receding/preceding motion along the line-of-sight naturally imprint a non-zero component to the $\mu_r(R)$ profiles. As outlined in \cite[van de Ven et al. (2006)]{vandeVen2006} and \cite[Bianchini et al. (2018)]{Bianchini2018} (Section 4.2) the perspective expansion/contraction is
\begin{equation}
\mu_r=-6.1363\times10^{-5}\, V_\mathrm{los}\,R/d \quad \mathrm{mas\,yr^{-1}},
\label{eq:expansion}
\end{equation}
with $V_\mathrm{los}$ the line-of-sight velocity of the GC, $d$ the distance to the cluster and $R$ the distance from the GC centre.
Any observed deviations from equation \ref{eq:expansion} can be taken as the signature of the presence of data systematics in \textit{Gaia} DR2.

An analysis of the closest 14 GCs (\cite{Bianchini2019}) shows that \textit{Gaia} DR2 performs excellently in recovering the expected radial proper motion profiles, indicating that systematics are below the random measurement errors, reaching a precision of $\sim0.015$~mas yr$^{-1}$ for mean proper motion measurements. In Figure \ref{fig1}, we illustrate the case of NGC~3201 that shows a consistency between observations and expectations within $<2-$sigma level.

\section{Tangential component of proper motions}
\label{sec:tangential}
\begin{figure}[b]
\begin{center}
 \includegraphics[width=0.9\textwidth]{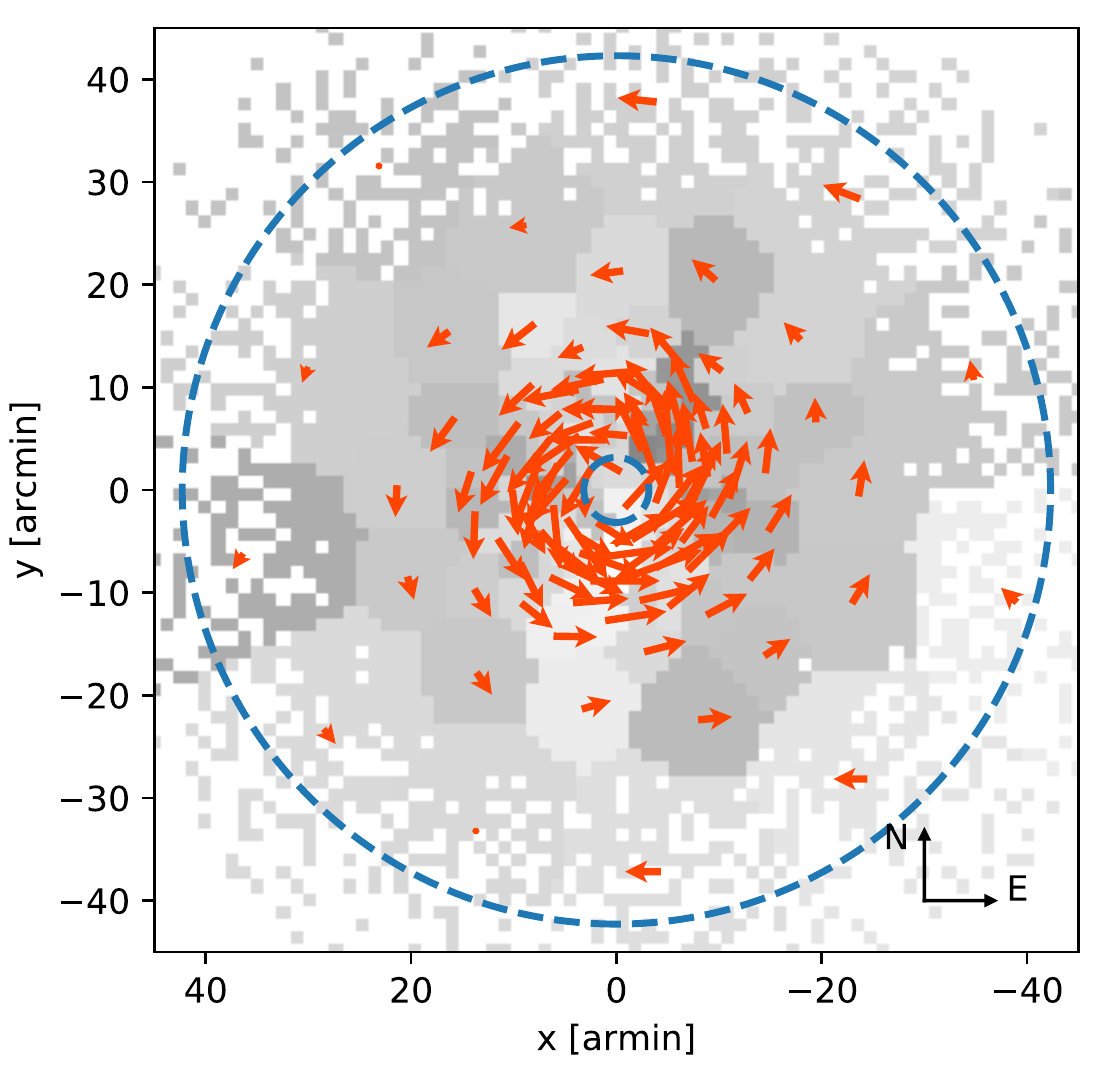} 
 \caption{\textbf{Velocity field of NGC 104}. The arrows indicate the proper motion vectors across the field of view of the cluster revealing a strong counter-clock rotation ($V/\sigma\approx0.50$). The dashed circles are the half-light radius and tidal radius of the cluster.}
   \label{fig2}
\end{center}
\end{figure}

The tangential component of proper motions $\mu_t(R)$ can be directly used to measure the presence of internal rotation on the plane of the sky for a large sample of GCs. Before \textit{Gaia} DR2, this was only possible using \textit{HST} proper motions available for a handful of GCs for which an absolute reference frame could have been built (e.g. using background stars belonging to the Small Magellanic Cloud, \cite{Bellini2017}).

We conducted a systematic search of internal rotation in 51 GCs, revealing that 11 GCs possess a mean tangential component of proper motions significantly different than zero within $>3-$sigma level, and within $2-$sigma level for other 11 GCs. In Figure \ref{fig2}, we report the velocity field-of-view for the GC NGC~104, clearly showing a strong presence of internal rotation. About $\sim50\%$ of GCs in our sample are observed to be rotating. Our results are consistent and complementary with previous line-of-sight studies (e.g. \cite{Kamann2018}) and \textit{Gaia} DR2 studies (\cite{Helmi2018,Sollima2019,Vasiliev2018}).

Finally, we quantify the amount of rotational support using the parameter $V/\sigma$, with $V$ the peak of the velocity rotation curve and $\sigma$ the central velocity dispersion, finding values of $V/\sigma\sim0.08-0.51$. Interestingly, the $V/\sigma$ distribution shows a correlation with both the relaxation time of the clusters and the GCs mass: GCs with longer relaxation times (i.e. the more massive GCs) display stronger internal rotation. Since angular momentum is expected to be dissipated during the long-term dynamical evolution of a GC (\cite{Tiongco2017}), these correlations indicate that GCs with longer relaxation times were more efficient in retaining angular momentum throughout their $>10$ Gyr long evolution. This finding suggests that internal rotation was a crucial ingredient already imprinted during GCs formation, and places some massive GCs (e.g. NGC 104) in the rotation regime typical of nuclear star clusters (\cite{Tsatsi2017}).

\section{Conclusions}
We presented an initial exploitation of \textit{Gaia} DR2 data, focusing in particular on the synergetic assessment of data systematics and the exploration of the internal dynamics of Milky Way GCs. Our results give the first taste of the unparalleled power of \textit{Gaia} DR2 for the understanding of GCs formation and evolution. A wealth of applications of the kinematic data still remains unexplored, but we anticipate that the detailed study of data quality -- in preparation for the subsequent data releases -- will provide a better understanding of the crowded GCs regions and unlock a number of physical properties hidden below the $\sim$km s$^{-1}$ regime.

\end{document}